\documentclass[11pt]{article}
\newcommand{\etal}{{\it et al.~\/}}

\newcommand{\lasr}{La$_{1-x}$Sr$_{x}$MnO$_3$~\/ }
\newcommand{\laca}{La$_{1-x}$Ca$_{x}$MnO$_3$~\/ }
\newcommand{\lcmo}{La$_{0.77}$Ca$_{0.23}$MnO$_3$~\/ }
\newcommand{\ndsra}{Nd$_{0.5}$Sr$_{0.5}$MnO$_3$~\/ }
\newcommand{\ndsrb}{Nd$_{0.45}$Sr$_{0.55}$MnO$_3$~\/ }
\newcommand{\prca}{Pr$_{0.6}$Ca$_{0.4}$MnO$_3$~\/ }
\newcommand{\ndca}{Nd$_{0.5}$Ca$_{0.5}$MnO$_3$~\/ }

\newcommand{\tc}{T$_c$~\/}
\newcommand{\tn}{T$_N$~\/}

\begin{document}
\baselineskip=20pt
\begin{center} {\Large {\bf An Electron Paramagnetic Resonance Study of  Phase Segregation  in  Nd$_{0.5}$Sr$_{0.5}$MnO$_3$  }}\\ 
\vskip 1cm
{\large Janhavi P. Joshi, A. K. Sood and S. V. Bhat \footnote{Corresponding author, {\it Email address:} svbhat@physics.iisc.ernet.in}}\\
{\it
Department of Physics,
Indian Institute of Science,
Bangalore 560 012, India
}\\
{\large Sachin Parashar, A. R. Raju and C. N. R. Rao}\\
{\it
Chemistry and Physics of Materials Unit, Jawaharlal Nehru Centre for Advanced Scientific Research, Jakkur P.O, Bangalore 560 064, India}
\end{center}
\vskip 0.5cm
\hrule
%\begin{abstract}
%\baselineskip=20pt

\vskip 0.2cm
\noindent{\bf Abstract}
\vskip 0.5cm 
\noindent We present results of an electron paramagnetic resonance (EPR) study of  Nd$_{1-x}$Sr$_x$MnO$_3$ with x = 0.5  across the paramagnetic to ferromagnetic, insulator to metal transition  at 260 K (T$_c$) and the antiferromagnetic, charge ordering transition   (T$_N$ = T$_{co}$) at 150 K. The results are compared with those on Nd$_{0.45}$Sr$_{0.55}$MnO$_3$ which undergoes a transition to a homogeneous A-type antiferromagnetic phase at T$_N$ = 230 K and on \lcmo which undergoes a transition to coexisting ferromagnetic metallic and ferromagnetic insulating phases. For x = 0.5, the EPR signals below T$_c$ consist of two Lorentzian components attributable to the coexistence of two phases. From the analysis of the temperature dependence of the resonant fields and intensities, we conclude that in 
the mixed phase ferromagnetic and A-type antiferromagnetic (AFM) phases coexist.  The x = 0.55 compound shows a single Lorentzian throughout the temperature range. The signal persists for a few degrees below \tn.  The behaviour of the A-type AFM phase is contrasted with that of the two ferromagnetic phases present in \lcmo. The comparison of behaviour of A-type AFM signal observed in both \ndsra and \ndsrb with the two FM phases of \lcmo,  vis-a-vis the shift of  resonances with respect to the paramagnetic phases and the behaviour of EPR intensity as a function of temperature conclusively prove that the \ndsra undergoes phase separation into A-type AFM and FM phases.


\vskip 0.2cm
\hrule
\vskip 0.5cm
\noindent {\bf 1 Introduction}
\vskip 0.5cm
Manganites of the general formula Re$_{1-x}$A$_x$MnO$_3$ where Re is a trivalent rare earth ion and A is a divalent alkaline earth ion, exhibit a number of interesting phenomena such as colossal magnetoresistance (CMR), insulator to metal, paramagnetic (PM) to ferromagnetic(FM) or antiferromagnetic(AFM) transitions and charge/orbitally ordered states as a function of x and temperature T \cite{rao,tok}. Not surprisingly, they have been the subject of intense study in the last few years. Recent  experimental and theoretical studies indicate that manganites undergo intrinsic phase separation (PS) in the CMR regime \cite{dag}. PS refers to the spontaneous and competing coexistence of two (or more) different phases such as FM clusters in an AFM background observed in the x-T phase diagram for x $\le$ 0.5, of large and intermediate electron bandwidth manganites. A large number of experimental techniques including transport,\cite {oga} magnetic, electron microscopy,\cite {ueh} scanning tunneling microscopy,\cite {fat} small angle neutron scattering, \cite {ter1} Brillouin scattering, \cite {mur} NMR \cite {all} and muon spin relaxation measurements,\cite {ter2,ter3} mainly on the prototype manganite La$_{1-x}$Ca$_x$MnO$_3$ have provided evidence for microscopically inhomogeneous FM  phases below the Curie temperature (T$_c$).  Recently the Nd$_{1-x}$Sr$_x$MnO$_3$ compound for x = 0.5 is also shown to spontaneously phase segregate \cite{woo,kaj,fuk}.  Woodward \etal\cite{woo} using  techniques such as x-ray diffraction and neutron diffraction showed a number of interesting features in its phase diagram. It was found that on cooling below T$_c$, an intermediate A-type AFM phase appeared in addition to the FM phase. Below T$_N$ = 150 K, the FM state  transforms to a CE-type AFM state in the temperature range of 150 K to 100 K, indicating the presence of FM phase in this temperature range apart from the A-AFM and CE-AFM phases. A study by Ritter \etal\cite{rit} also illustrates coexistence of these three phases at 125 K. It was suggested that such phase segregation aids the field induced structural transitions observed in \ndsra  (NSMO0.5) \cite{mah}.   While most of these studies report the occurrence  of PS below T$_c$,  photoemission studies indicated the presence of a phase separated state even above T$_c$ \cite {dag}.

Theoretical attempts to understand PS have been mainly through numerical simulations and mean - field calculations. Two different scenarios have emerged, {\it viz} electronic phase segregation and disorder-induced mixed state. In the former, a phase separated regime interpolates between the AFM and FM phases while approaching the FM phase by hole doping of the AFM phase.   When long-range Coulomb interactions are taken into account, the phase separated domains turn out to be of the order of a few nanometer size as observed in some experiments \cite{kir}.
However, experimental results of Uehra \etal \cite{ueh} indicate the presence of much larger ($\sim$ micrometer size) domains. It is well known that the A-site cation radius $<r_A>$ plays a major role in determining the properties of manganites. It was also found \cite{rod,mor} that the disorder caused by mismatch in  the ionic sizes could  play a significant role in phase separation as well as in determining the CMR properties.  The size of the A-site cation influences the Mn-O-Mn bond angle which in turn affects the hopping rate of the charge carriers. A random distribution in the size of A-site cations leads to a random distribution of the rate of hopping and  the random distribution of the exchange coupling constant J which is also determined by the Mn-O-Mn bond angle. A local variation in these parameters, leads the system to be either in FM or AFM state depending on the parameters. On the other hand, creating a large number of FM-AFM interfaces is not energetically favourable. Competition between these two tendencies in a system leads to the formation of clusters much larger than the lattice spacing in which the number of interfaces is reduced and the tendency for local disorder to form either a FM or an AFM phase is partially satisfied. A homogeneous phase would therefore be expected for a compound such as PrCaMnO$_3$ where the A-site cation radii are comparable. As the size disorder grows,  the cluster size goes on reducing in the phase separated state. Looking at the ionic radii of various A-site cations in 6-coordinated octahedral structures 
 (Nd : 1.123 A$^0$, Pr: 1.13  A$^0$, Ca : 1.14  A$^0$, Sr : 1.32  A$^0$)
one can expect that NdCa or PrCa manganites have more homogeneous phases than NdSr or PrSr manganites which are expected to phase segregate as  indeed  found experimentally \cite {kaj,zvu}. In spite of a large number of experimental and theoretical studies, a clear and consistent picture of the PS phenomenon is yet to emerge. For example the recent X- band EPR study of Rivadulla \etal \cite {riv} proposes a coexistence of a FM phase with a PM phase in LaCaMnO$_3$ instead of the FM and AFM phases as believed commonly.

Since EPR is a very powerful local probe and is found to be sensitive to the presence of inhomogeneities, we have used the technique to study
 Nd$_{1-x}$Sr$_x$MnO$_3$ single crystals and powders with a view to investigating the phenomenon of  phase separation . We have used  samples with x = 0.5 (NSMO0.5) and x = 0.55 (NSMO0.55). The phase diagram of x = 0.5 sample is fairly complex \cite {kaj} with coexisting phases in certain regions of the phase diagram. It is a paramagnetic insulator at room temperature. Neutron diffraction studies, however, reveal presence of two-dimensional FM fluctuations \cite {kaw} in the paramagnetic state in the vicinity of T$_c$  = 260 K  where it transforms to a FM phase. Below  200 K,  A-type AFM peaks in the neutron diffraction experiments are observed along with the FM phase indicating a coexistence of FM and A-type AFM phases \cite {kaj}. The A-type AFM phase consists of FM planes coupled antiferromagnetically to each other and is characterised by the d$_{x^2 - y^2}$ type orbital ordering. Below the Neel temperature T$_N$ = 150 K a CE-type AFM state is formed which also coexists with the A-type AFM state \cite {kaj}.  The CE-type state is characterised by 3x$^2$ - r$^2$/3y$^2$-r$^2$ orbital ordering. The Neel temperature is also the charge ordering  temperature in this material. 
NSMO0.55 on the other hand is not known to show any phase separation. It is a paramagnetic insulator at room temperature  and transforms to an  A-type AFM state at T$_N$ = 230 K with    
d$_{x^2 - y^2}$ orbital ordering without any charge ordering \cite {kaj}. We compare the EPR results of NSMO0.55 which does not show any phase segregation with those of NSMO0.5 and find that these two samples behave very differently \cite{jan1}. 
To compare and contrast with the results on these two materials we have also studied \lcmo (LCMO). LCMO (\tc $\sim$ 230 K) shows phase separation into two ferromagnetic phases, one metallic and the other insulating \cite{pap}. Thus we can expect very different behaviour of various lineshape parameters in LCMO from the ones in the \ndsra and \ndsrb.
\vskip 0.5cm
 \noindent {\bf 2 Experimental}
\vskip 0.5cm
The single crystals of NSMO0.5 were prepared by the float zone technique.  The resistivity measurements show a transition from  insulating state to a metallic state at T$_c$ = 260 K.  Further transition to an AFM insulating state is observed at T$_N$ = 150 K. Magnetization measurements show a sharp increase in the susceptibility at 260 K, which remains practically temperature independent  till $\sim$ 150 K, below which it shows a sharp drop.  Single crystals of NSMO0.55 were also prepared by the same method. The resistivity measurements show a monotonic increase with decreasing temperature and a sharp increase at T$_N$ = 230 K.  The magnetization measurements show a monotonic increase in the susceptibility till 230 K with a sharp fall at this temperature.

 The EPR experiments were carried out on both single crystal and powder samples of NSMO0.5 using a Bruker X - band spectrometer (model 200D) equipped with an Oxford Instruments continuous flow cryostat (model ESR 900). Similar experiments were carried out on a single crystal and powdered single crystal of NSMO0.55 and  LCMO. The spectrometer was modified by connecting the X and Y inputs of the chart recorder to a 12 bit A/D converter which in turn is connected  to a PC enabling digital data acquisition. With this accessory, for the scan-width typically used for our experiments i.e 6000 Gauss, one could determine the magnetic field to a precision of $\sim$ 3 Gauss. For single crystal study the static magnetic field was kept parallel to the c-axis of the crystals. The temperature was varied from 4.2 K to room temperature (accuracy: $\pm 1 K$) and the EPR spectra were recorded while warming the sample. For measurements on powder the samples  obtained by finely crushing the single crystals of the respective materials ,   were dispersed in paraffin wax. During all the experiments a speck of DPPH  was used as a g-marker whose signal was subsequently subtracted digitally to facilitate lineshape fitting.
\vskip 0.5cm
\noindent {\bf 3 Results and Discussion}
\vskip 0.5cm
Figure 1 shows a few typical EPR signals of those recorded from  the powder sample of NSMO0.5 in the temperature range from 290 K to 80 K. The signal intensity goes on decreasing below 150 K and below 80 K the signal becomes too weak to analyse. As can be seen from the figure the relatively narrow EPR lines in the PM phase become  broad below T$_c$. Further they show considerable variation in the lineshapes  as a function of temperature. Based on the nature of the lineshapes, we have divided the whole temperature range into five different regions and have fitted appropriate lineshape functions to the signals in each region. In figure 1 we have indicated the region to which the respective signals belong. The method of analysis followed for each region  is described below. 

\begin{enumerate}
\item In the temperature range between 290 K and 255 K (the signals at 290 K and 277 K in figure 1), the signals are narrow and are well fitted to the derivative of a single  Lorentzian described by equation: \cite{iva}  
$$ {dP\over dH} = {d\over dH}({\Delta H\over (H-H_0)^2 + \Delta H^2})\eqno(1) $$
where $\Delta$ H is the linewidth and H$_0$ is the center field.

\item  Distorted signals  are observed in the temperature range of 250 K to 200 K as shown by data for 210 K. These signals are also quite broad and therefore they have to be fitted with the derivative of Lorentzian consisting of two terms given by equation 

$$ {dP\over dH} = {d\over dH}({\Delta H\over (H-H_0)^2 + \Delta H^2}+ {H\over (H+H_0)^2 + \Delta H^2})\eqno(2)$$ 

where the first term represents signal response due to the clockwise polarized component of microwaves and the second term represents the signal due to the anticlockwise polarized component of the microwaves \cite {iva}. In this temperature range, we found that we could obtain better fits by using two Lorentzian functions with different linewidths and center fields, each described by equation 2.
  
\item Much broader signals are observed in the temperature range from 200 K to 155 K which could be very well fitted to the lineshape function as described in (2) above consisting of two Lorentzians (signal at 180 K shown in figure 1). 

The separation into  regions  two and three has been made based on the quality of fits which we have obtained using the same equation in the two regions. It can be clearly observed that the quality of fits continuously goes on improving as the temperature is reduced in region 2 till 200 K (note the difference between the experimental curve and the fitted curve at 240 K and 210 K), indicating an evolution of lineshapes which results in the lineshapes of region 3.

\item In the temperature range between 150 K to 100 K we observe the development of a distortion in the signal close to zero field (indicated by a  circle on the signal at 140 K in figure 1). The signals in this range could not be fitted either to a single Lorentzian as in region 1 or to two Lorentzians as in regions 2 and 3. However if the data set in the low field region is excluded we could fit these signals to a single Lorentzian.
 
\item  Undistorted broad signals between 100 K and 80 K are observed which fit excellently to a single Lorentzian (equation 2)(signal at 80 K shown in figure 1).
\end{enumerate}

In Figures 2a and 2b we plot the center fields (H$_0$) and linewidths obtained from the lineshape fits as described above as a function of temperature. The vertical dotted lines show the different regions in which we categorized  the different  temperature ranges from 80 K to 290 K. As can be seen from  Figure 2a, the center field at 290 K  is   3407 Gauss ($\pm$ 3 Gauss) corresponding to `g'=1.976 which is less than the free electron `g' value. 
In region 2 where we have fitted the signal to two Lorentzians, H$_0$ of one of the components is less than  H$_0$ in  region 1 (component `P'). The H$_0$ of this component goes on decreasing with decreasing temperature upto 200 K. The H$_0$ of the second component (component `Q') on the other hand is greater than the H$_0$ in  region 1 and it shows a weak temperature dependence and increases as the temperature is reduced. In the third region these two components have opposite temperature dependences to that in region 2. The `P' component shows an increase in H$_0$  as the temperature is reduced remaining less than the H$_0$ value in region 1. Component `Q' on the other hand shows a decrease as a function of decreasing temperature. In region 4 the single H$_0$ of the Lorentzian fitted to  the signal excluding the low field region, shows a non-monotonic behaviour and has a magnitude less than that in  region 1. In region 5, H$_0$ is practically temperature independent. 

As can be seen from Fig. 2b, the linewidth of the single Lorentzian fitted to the signals in region 1 shows an increase as a function of decreasing temperature. This increase continues in regions 2 and 3 for both the Lorentzian components `P' and `Q' fitted in these regions. In region 3 at a temperature 170 K $\sim $1.1 T$_N$  a peak in the linewidth is observed. Below this temperature in region 3 linewidths of both the signal components decrease with decreasing temperature. In regions 4 and 5 it again increases as a function of decreasing temperature. It is noteworthy  that unlike the other manganites studied before \cite{raj,jan} signals do not disappear in the antiferromagnetic phase i.e below T$_N$ = 150 K. 

%According to Ritter \etal ~the system may be consisting of three different phases: a CE-type AFM, an A-type AFM and a FM phase.
%However, looking at the EPR intensity of the signal components observed in the three compounds, we can conclude that the signals in region 4 and 5 may be attributed to the FM phase.

 %Therefore, the EPR signal observed in this temperature range, on the basis of EPR results on NSMO0.5 alone could  be attributed to either the A-type AFM phase or the FM phase present in this temperature range. 
%However, a comparison with our results on NSMO0.55 enables us to show that these signals are from the A-type AFM phase.  

Figure 3 shows the signals from the single crystal of NSMO0.5 at different temperatures. At temperatures from 290 K to 260 K the signals have Dysonian lineshapes which progressively distort as the temperature is lowered below 260 K and evolve into very broad lines presumably consisting of more than one signal. A distortion near the zero field similar to  signals in the powder sample appears in these signals as well below 150 K. The line further broadens, decreases in intensity as the  temperature is reduced below 100 K. Below $\sim$ 80 K the signal becomes very broad with a signal to noise ratio too poor for analysis.  The signals down to 270 K could be fitted to the Dysonian lineshape \cite {iva}.

$$ {dP\over dH} = {d\over dH}{\Delta H + \alpha (H-H_0)\over (H-H_0)^2 + \Delta H^2}\eqno(3) $$

where $\alpha$ is the asymmetry parameter representing the ratio of the absorption component  to the dispersion component of the response. However it was not possible to fit the signals below that temperature to any of the following models: (i) single Lorentzian with two terms, to take into account both left and right circularly polarised microwave fields. (ii) two Lorentzians (accounting for phase segregation) with  two terms each to take into account both left and right circularly polarised microwave fields. (iii) single Dysonian with two terms to take into account both left and right circularly polarised microwave fields. (iv) two Dysonians (accounting for phase segregation) with two terms each to take into account both left and right circularly polarised microwave fields. 
Keeping in mind the fact that the signals from  the powder of the same sample could be adequately fitted, we conclude that the inability to do so in case of the single crystal samples is due to the effect of the multidomain structure which may be present. 

In the inset of figure 4 we show few representative signals obtained from powder sample of LCMO. The resonance fields obtained by fitting these signals (to a single Lorentzian described by eq. 1 in the paramagnetic phase and to two Lorentzians, each described by eq. 2, below 165 K) are also shown in the figure.  

Figure 5 shows the signals from the powder sample of NSMO0.55 from 200 K to 290 K. Below 200 K the signal is too weak to analyse. The signals are narrow and are symmetric. They are fitted to the   Lorentzian lineshape  and the fitting parameters thus obtained, {\it viz},  H$_0$, peak to peak linewidth $\Delta$H$_{pp}$ are  plotted in Fig. 6 as a function of temperature. As shown in Fig. 6a the linewidth monotonically decreases as the temperature is reduced till 210 K below which it shows a sharp increase. The minimum in the linewidth occurs at 210 K. Here again, we observe a signal in the AFM state which is A-type. The resonance fields H$_0$ plotted in fig. 6b  show a weak temperature dependence throughout the temperature range increasing till 210 K and then shows a decrease. The g value obtained from the resonance fields is less than the free electron g value and remains so throughout the temperature range.  The intensity  shows a peak at 210 K and then decreases sharply. It roughly follows the temperature dependence  shown by dc susceptibility.

In figure 7, we plot the double integrated intensities of different signal components in the three materials for the sake of comparison. 

Based on an analysis of the different lineshapes of the powder signals of NSMO0.5 we make the following observations. 
In region 1, the signals fit very well to the derivative of a single Lorentzian. Clearly, the sample is monophasic and homogeneous. Neutron diffraction measurements \cite {woo,kaj}  also show that the sample is paramagnetic and monophasic in this region. The `g' value obtained from the EPR experiments is smaller than the free electron `g' value as expected for a CMR manganite in contrast with the hole doped CO manganites where `g' values are greater than the free electron values.  Linewidths in this region are much smaller than those of the CO manganites studied by us earlier  in their paramagnetic state \cite{raj,jan}.  
As we go into region 2, the quality of the fits to the derivative of single Lorentzian deteriorates. This may be understood in terms of the building up of  FM correlations as verified by neutron diffraction \cite {kaw} and by the two-dimensional nanoscopic structural correlations  as proposed by Kiryukhin  \etal \cite {kir} based on x-ray diffraction experiments. Presence of such fluctuating FM correlations may cause  distortion of a Lorentzian signal due to the fluctuating internal fields. We have fitted these signals to the derivative of Lorentzians containing two components. The center fields corresponding to these two components have magnitudes shifted above and below the center field in the PM region.  We attribute this behaviour to the presence of  A-type AFM (component Q) and FM (component P) phases in the sample, respectively.  Detailed evidence will be provided below in support of this assignment. 

Now we would like to consider  another possibility which could possibly give rise to a multiline spectrum in the ferromagnetic state. It has been observed in  single crystal specimen of ferromagnetic metals like Nickel \cite{bhagat} and Iron \cite{kip} that a multiline complex spectrum results due to the magnetic anisotropy of the ferromagnetic sample. When the applied magnetic field is oriented away from the easy axis, the two components of magnetisation, one driven by the torque of the anisotropy field and the other due to the applied magnetic field, can cause the splitting of the EPR spectrum. Similar splitting was observed in a polycrystalline ferrite sample by Schlomann \etal \cite {sch} who have studied the temperature dependence of the  resulting EPR spectrum.  This happens when the anisotropy fields are large with respect to the resonant field. However the situation in manganites is different. A study of the effect of magnetic anisotropy on EPR signals of manganites  has been carried out by Lofland \etal \cite {lof10} in single crystalline \lasr  and  they estimate the anisotropy field  to be  $\sim$ 160 Gauss. The anisotropy field in NSMO0.5 could not be very different from that in \lasr since the two are structurally similar high bandwidth manganites. Therefore, the anisotropy field in NSMO0.5 is much less than the resonance field which is $\sim$ 3000 Gauss.

 Schlomann \etal \cite{sch} have done the temperature study of split signals in ferrites. They observe that due to magnetic anisotropy of their ferrite samples, the splitting of their signals increases as the temperature is decreased. However, both the components of the signal shift towards the low field side as a function of decreasing temperature and the intensity of the low field component continuously increases. Our observations in  NSMO0.5 are qualitatively very different from these as seen from Fig. 2a. The intensity of the two signal components also does not follow the trend as seen by Schlomann {\it et al.} Moreover, the two line structure observed in the polycrystalline samples is due to a powder pattern \cite{rai} and does not fit to two Lorentzian lineshape functions. Because of these reasons we believe that the observed splitting of the EPR signal cannot be attributed to the magnetic anisotropy in NSMO0.5. In a recent work Shames \etal \cite{shames2} derive similar conclusion for their EPR study  on \laca (x=0.5)

Another complicating factor in magnetically ordered materials, namely,  the demagnetizing fields due to possibly different  sized and shaped particles does not seem to be significant in  our experiments. As mentioned by Lofland \etal \cite{lof10} and Patil \etal \cite{patil} demagnetizing fields cause a  shift in the position of the DPPH signal. Throughout the temperature range, we did not observe any change in the DPPH position, even across \tc  when the demagnetizing factors were expected to come into picture. 

  As seen in Fig. 2b, the linewidths in the  regions 2 and 3 are  much larger compared to those in CO manganites probably due to the large distribution  of internal fields in the FM state. 
The linewidths of the AFM component are also large as expected, and are comparable to linewidths of the FM phase. A peak in the linewidth in region 3 appears at nearly 1.1T$_N$.  In region 4, only CE-type and  A-type phases are expected to be present according to the neutron diffraction studies of Kajimoto \etal \cite {kaj}.   However the signals in this region show distortion near the zero field. This is the temperature range just below T$_N$. The lineshape has been fitted to the part of the signal excluding this distortion. The temperature dependence of the center field is non-monotonic showing a peak at 130 K. The linewidth also shows a non-monotonic temperature dependence in region 4, passing through a minimum at  130 K.  Below 100 K i.e in region 5 according to the neutron diffraction studies of Kajimoto \etal and Woodward \etal A-type and CE-type phases are present. The signal observed in this  region, a  broad but a single Lorentzian fits very well to equation 2. We can attribute this signal to the A-type AFM phase or the FM phase which gradually converts to CE phase since CE-type AFM phase does not give an EPR signal as shown by our previous studies in CO manganites \prca and \ndca \cite {raj,jan}. The linewidth in this region increases monotonically with center field being practically temperature independent. We shall come back to the origin of these signals in the subsequent discussion.

From the above analysis  of results of NSMO0.5 powder, it is clear that the   EPR lineshapes are very sensitive to the magnetic inhomogeneity of the sample. The analysis of the single crystal data was made difficult by large linewidths, large internal field effects and the sensitivity of the signals to the orientation of sample with respect to the magnetic field. Fitting of the highly distorted signals below T$_c$ was practically impossible. Below 100 K,  only a part of the line could be recorded in the available field range as can be seen from Fig. 3. 

In the previous discussion of the EPR results on NSMO0.5, we have presented two important results, (1) EPR presents evidence for phase segregation and (2) This phase segregation in NSMO0.5 involves, separation into a ferromagnetic phase and an A-type antiferromagnetic phase. In the following we provide additional evidence to support these conclusions.  We present EPR results on \lcmo (LCMO), which is known to phase separate into two ferromagnetic phases \cite{pap} and on \ndsrb (NSMO0.55) which is known to undergo a transition to a homogeneous A-type AFM phase.
\vskip 0.5cm
\noindent {\bf EPR in LCMO}
\vskip 0.5 cm
As can be seen from the inset of fig. 4, the EPR signals in LCMO from room temperature to 225  K (\tc $\sim$ 230 K) are narrow (denoted by symbol `A') and can be fitted to the Lorentzian lineshape function given by eqn 1. The signals in the range 225 K - 170 K (see for example the signal at 210 K)  seem to be consisting of two components.  The narrow signal component `A' observed in this temperature range represents the presence of  residual paramagnetic phase. The ferromagnetic phase gives rise to the broad signal component `B'. The component `A'  decreases in intensity with decreasing temperature and the component `B' increases in intensity resulting in a lineshape shown at 165 K. The signals from 165 K down to lowest temperature however could be very well fitted to two Lorentzians each represented by eqn 2. We have obtained the resonanant fields and the double integrated intensity from these fits.  The resonance fields plotted in fig. 4 are practically temperature independent in the paramagnetic phase. In the ferromagnetic phase however, both the signal components show a resonance field shifted to lower field side compared to that in the paramagnetic phase as expected from a ferromagnetic phase. This is consistent with the earlier evidence from NMR experiments \cite{pap} that the two phases  at this composition are ferromagnetic (one being insulating and another being metallic.) It is to be noted that we start observing the signatures of phase separation at a higher temperature than the NMR studies. We believe that this is because of the different timescales of the two techniques,  EPR being a much faster probe than NMR. 
A very important point to note here is that the shift of resonance fields in LCMO in the phase separated regime is very different from that of NSMO0.5, in that, one of the two signals (marked as `Q') in NSMO0.5 shifts to a resonanant field higher than the resonant field in the paramagnetic phase and the other signal (marked as `P') shifts to a lower field than the signal in the paramagnetic phase. This low field shift indicates that component `P' is obviously from the ferromagnetic phase present in the sample in this temperature range.
We propose that the high field signal component in region 2 and 3 in fig. 2a is indeed due to A-AFM phase which is present in the sample in that temperature range. To substantiate this claim we now discuss the results of NSMO0.55 which is a single phasic  homogeneous sample but goes into an A-AFM phase.
\vskip 0.5cm
\noindent {\bf EPR in NSMO0.55}
\vskip 0.5cm
Figure 6 presents the EPR parameters $\Delta$H$_{pp}$ and H$_0$ obtained from the Lorentzian fits to the signals of NSMO0.55.
 
Throughout the  temperature range in which we observe the EPR signal, we see that the signal is a single Lorentzian  as expected for a  homogeneous powder sample. It may be noted that we observe a signal below the Neel temperature unlike the CO manganites studied before \cite{raj,jan} but similar to NSMO0.5 which has an A-type AFM phase below 200 K. We would also like to mention here that the resonance field of the component Q of NSMO0.5 also has a `g' value smaller (as indicated by the shift of the signal towards higher fields) than the paramagnetic `g' value similar to NSMO0.55 signals. Further, the temperature dependence in the two cases is also qualitatively similar which supports our interpretation that the component Q of NSMO0.5 signals is due to the A-type AFM phase. 

Finally, we  analyse the intensities of the EPR signals in the three materials studied here. 
As can be seen from fig. 7a, in case of NSMO0.5, the EPR intensity in the PM phase shows a weak temperature dependence as expected for the PM phase. The intensity of both the signal components below \tc rises very sharply. However, below 200 K which is reported to be the Neel temperature of A-AFM phase \cite {woo,kaj,rit}, the intensity of the component Q drops very rapidly. A similar sharp drop of intensity  below 210 K is also seen in NSMO0.55 (fig. 7b). 
 On the other hand the intensity of the P component  shows a much broader peak indicating nearly temperature independent intensity down to 150 K.  The FM phase in the temperature range of 240 K to 150 gradually changes over to CE phase below 150 K \cite{woo}. Ritter \etal \cite{rit} have also observed the presence of FM phase at 125 K. Hence it is to be expected that the signal intensity of the component P drops gradually as the amount of FM phase in the sample decreases gradually. The resonance fields of these signals are below PM values. Hence we conclude that the behaviour in regions 4 and 5 of fig. 2a and 2b in the temperature range of 150 K to 80 K are to be attributed to the residual FM phase in the sample. 

In  marked contrast to this behaviour is the behaviour of intensity of the two phases in LCMO. (fig. 7c) The intensity of these two phases remains practically temperature independent down to very low temperature as is to be expected for any FM phase which remains stable down to very low temperature.

In conclusion, the present study of  x=0.5 and 0.55 compositions of Nd$_{1-x}$Sr$_x$MnO$_3$ and \lcmo demonstrates that EPR can be used as a powerful diagnostic tool for investigating phase segregation in manganites.  

While this paper was about to be submitted, we came to know of the work by Rivadulla \etal \cite{riv1} which also reports briefly on their EPR study of Nd$_{0.5}$Sr$_{0.5}$MnO$_3$. While both of our works report the presence of a mixed phase in this compound, Rivadulla \etal conclude that ferromagnetic and {\it paramagnetic} phases are present below T$_c$ while we believe for the reasons described in detail in this paper, the mixed phase consists of FM and {\it A-type antiferromagnetic} phases. 
\vskip 0.5cm
\noindent {4 \bf Acknowledgments}
\vskip 0.5cm
AKS and  SVB thank Department of Science and Technology and University Grant Comission for financial support. JJ thanks CSIR, India for a fellowship and F. Rivadulla for helpful critical comments.

\newpage
\noindent {\bf Figure Captions}\\

\noindent { FIGURE 1:}

 EPR spectra of a powder sample of Nd$_{0.5}$Sr$_{0.5}$MnO$_3$ for a few representative temperatures. The solid circles are experimental data and the solid line shows the fit to the appropriate equation as described in the text. The numbers on the left of each spectrum indicate the region to which it belongs. The region indicated by a circle at 140 K is excluded from fitting.\\ 
\noindent {FIGURE 2}

Temperature dependence of (a) the center fields H$_0$ and  (b) the linewidths $\Delta$H obtained from the fits to EPR spectra of powder \linebreak Nd$_{0.5}$Sr$_{0.5}$MnO$_3$.  The vertical dotted lines indicate different regions in which the temperature range is divided. The two signal components obtained by the fits are indicated by the letters P and Q in both Fig. a and b. \\
\noindent {FIGURE 3}

EPR spectra of a single crystal sample of Nd$_{0.5}$Sr$_{0.5}$MnO$_3$ at a few representative temperatures. The solid circles are the experimental data points. The smooth lines shown at some temperatures are fits to (i)a single Dysonian lineshape function (290 K and 270 K) and (ii) two Dysonians (245 K) (equation 3).\\
\noindent {FIGURE 4}

The resonance fields of \lcmo powder signals plotted as a function of temperature. The inset shows a few representative signals from \lcmo powder. The signal component denoted by `A' represents the signal from the  paramagnetic phase and the one denoted by `B' represents the signal from the ferromagnetic phase. The solid circles in the inset are the experimental data and the solid lines are the fits to the appropriate equations described in the text. \\
\noindent {FIGURE 5}

EPR spectra of a powder sample of Nd$_{0.45}$Sr$_{0.55}$MnO$_3$ at a few representative temperatures. The solid circles are the experimental data and the solid lines are fits to the Lorentzian lineshape (equation 3) .\\ 
\noindent {FIGURE 6}

Temperature variation of the lineshape parameters for the powder sample of Nd$_{0.45}$Sr$_{0.55}$MnO$_3$ (a)  linewidth(b)~H$_0$ obtained from the fits to  equation 1. The inset shows the g values calculated from H$_0$.\\
\noindent {FIGURE 7}

The normalised double integrated intensity (area under the curve) of the EPR signal components plotted as a function of temperature for (a) NSMO0.5 (b) NSMO0.55 (c) LCMO samples. The lines are guides to the eye.\\
\pagebreak

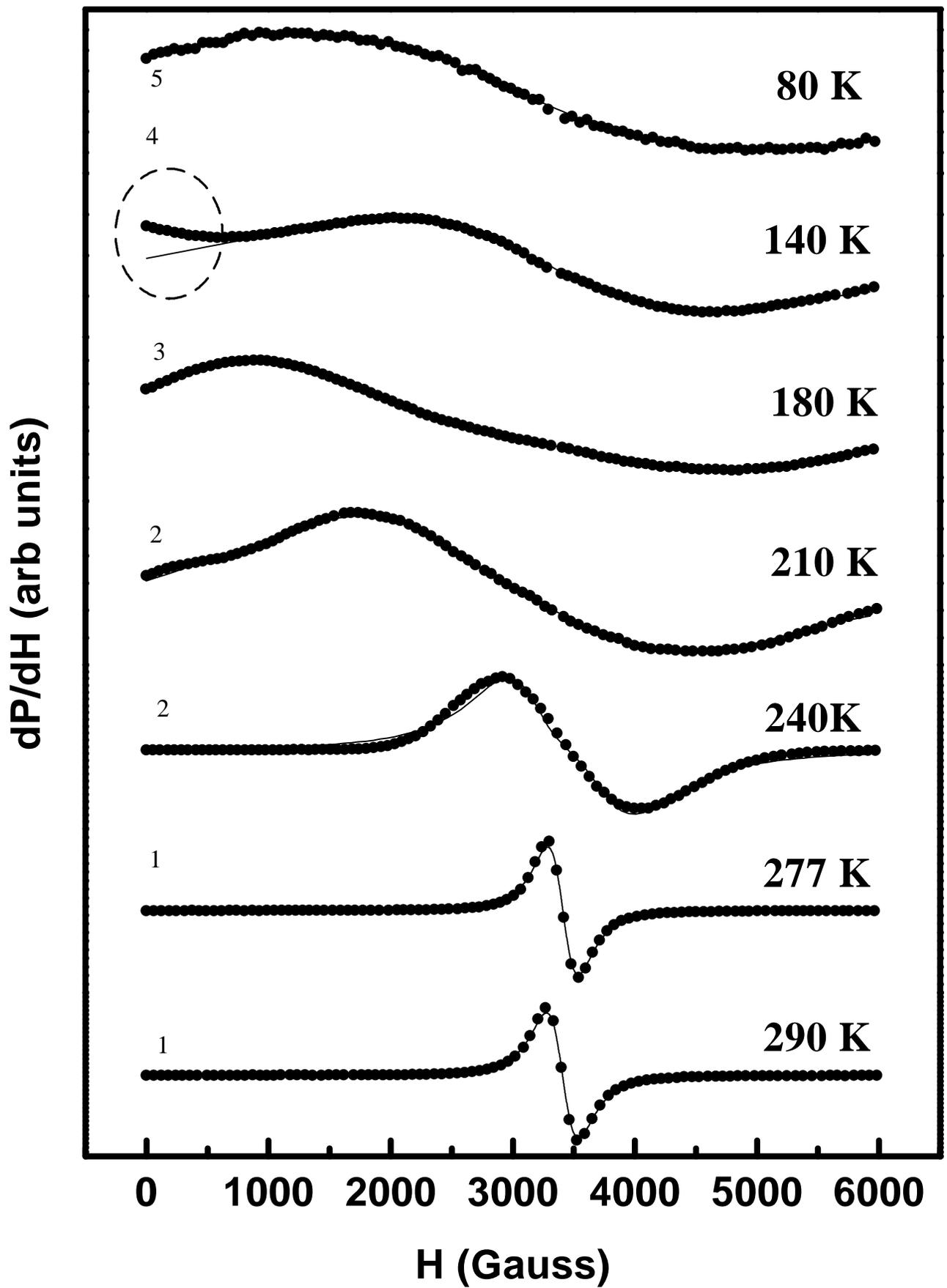
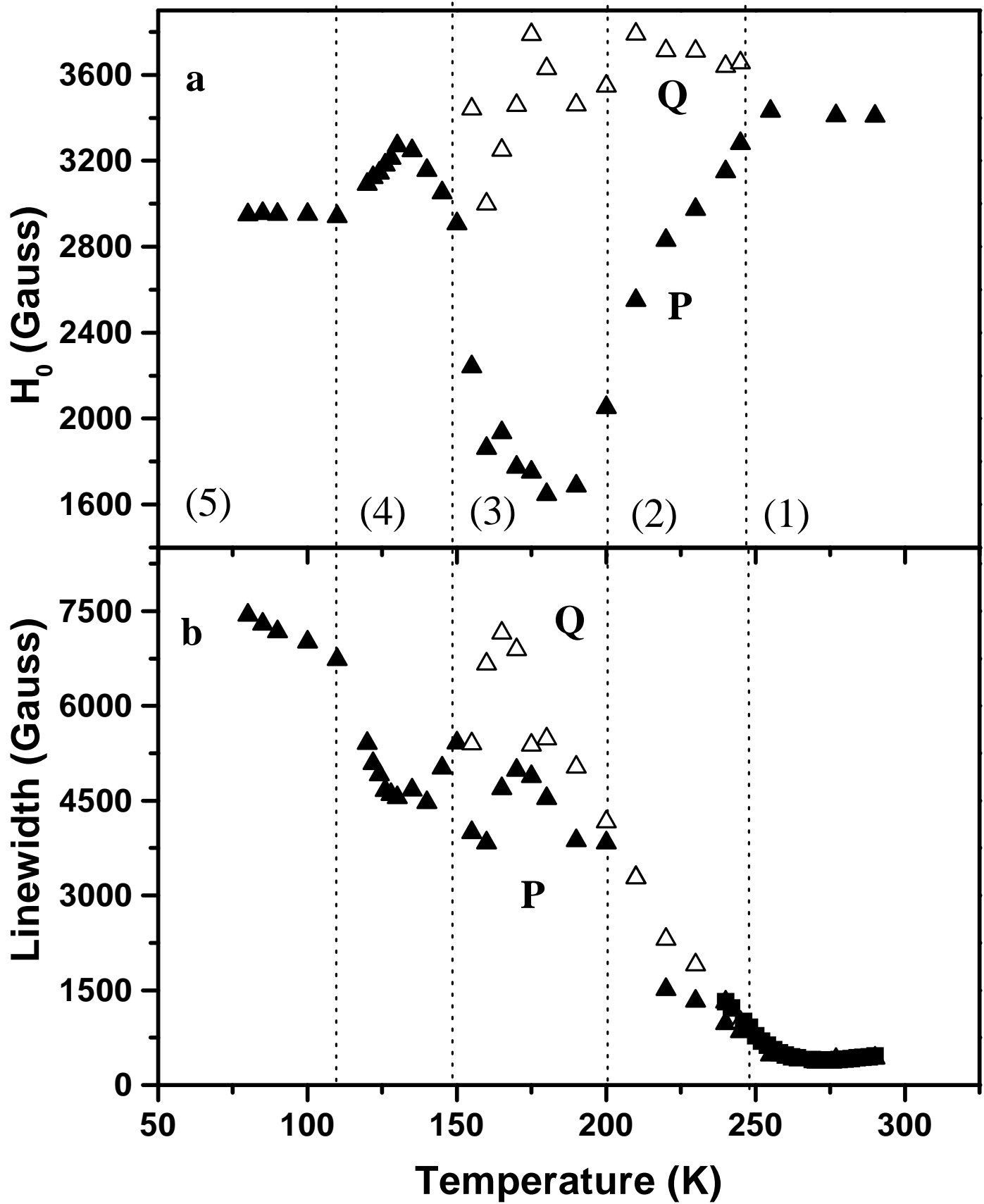
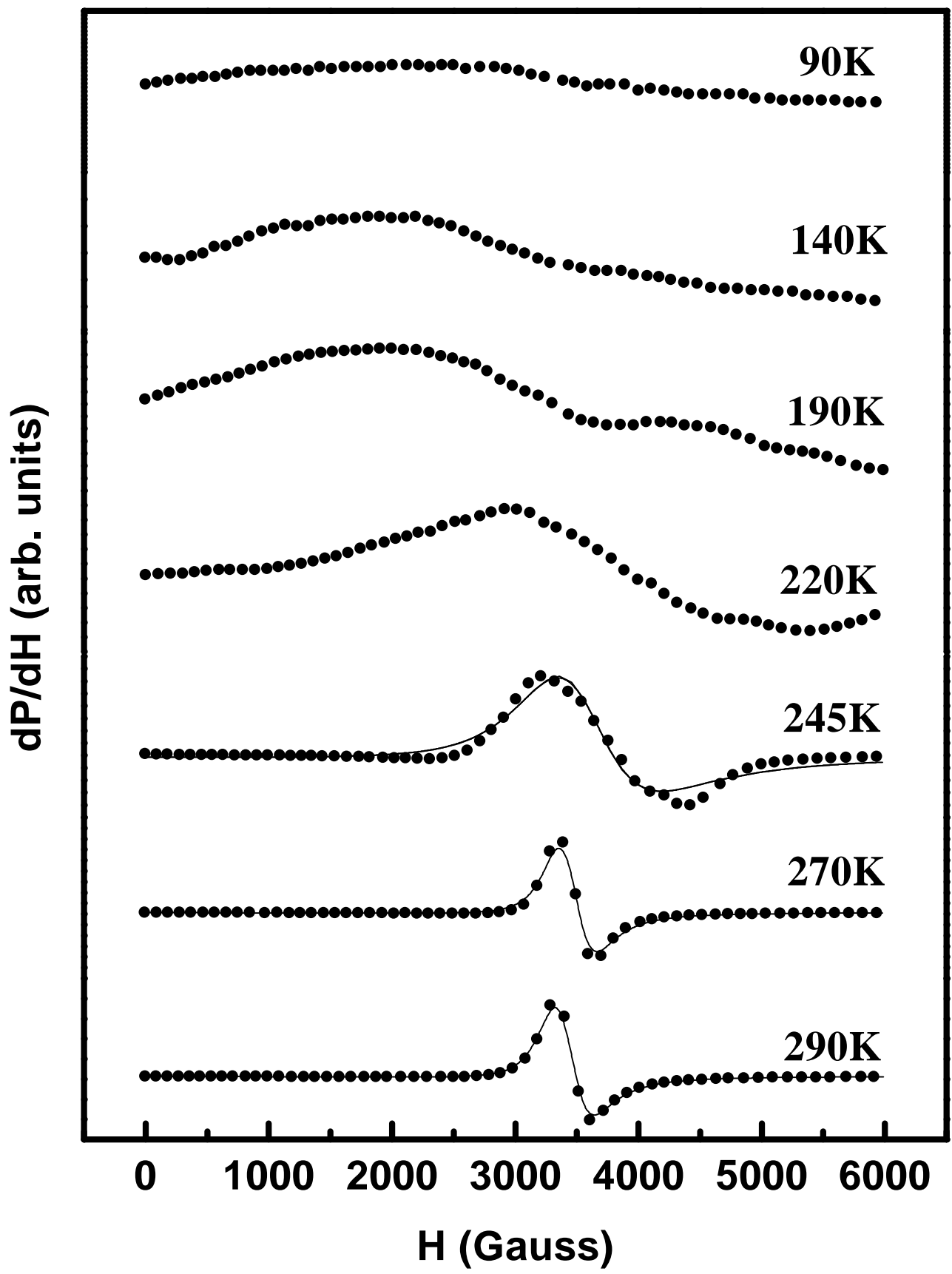
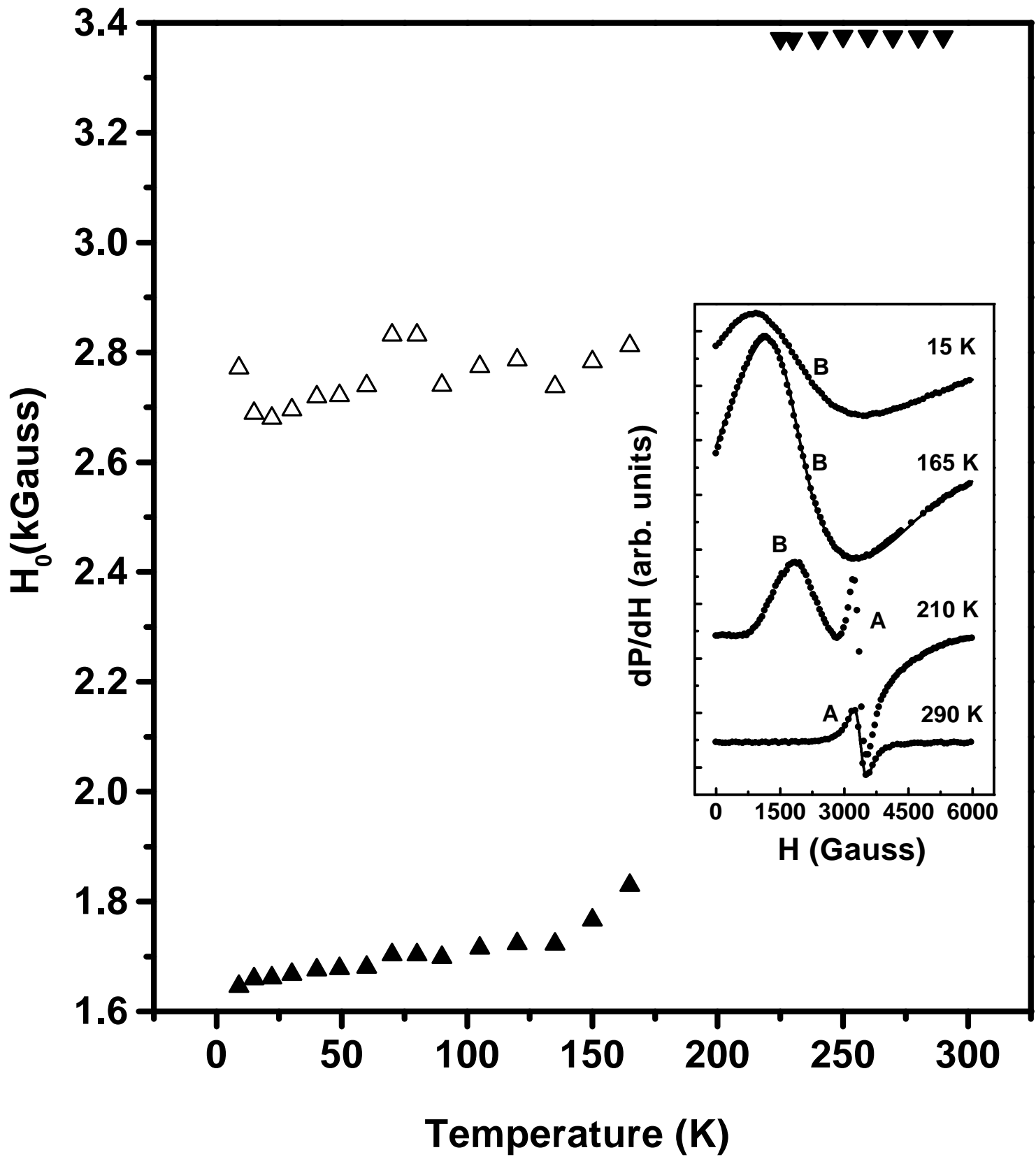
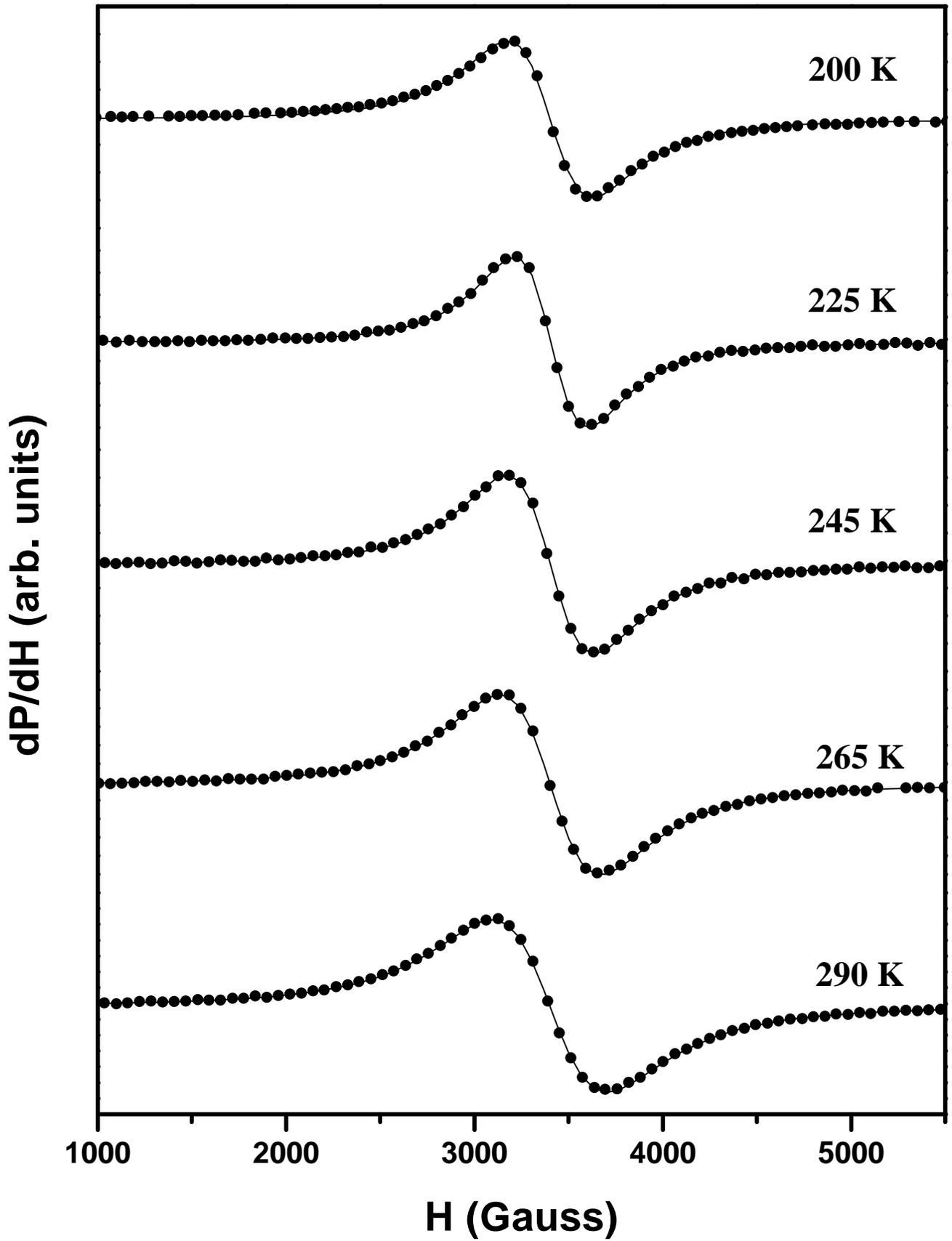
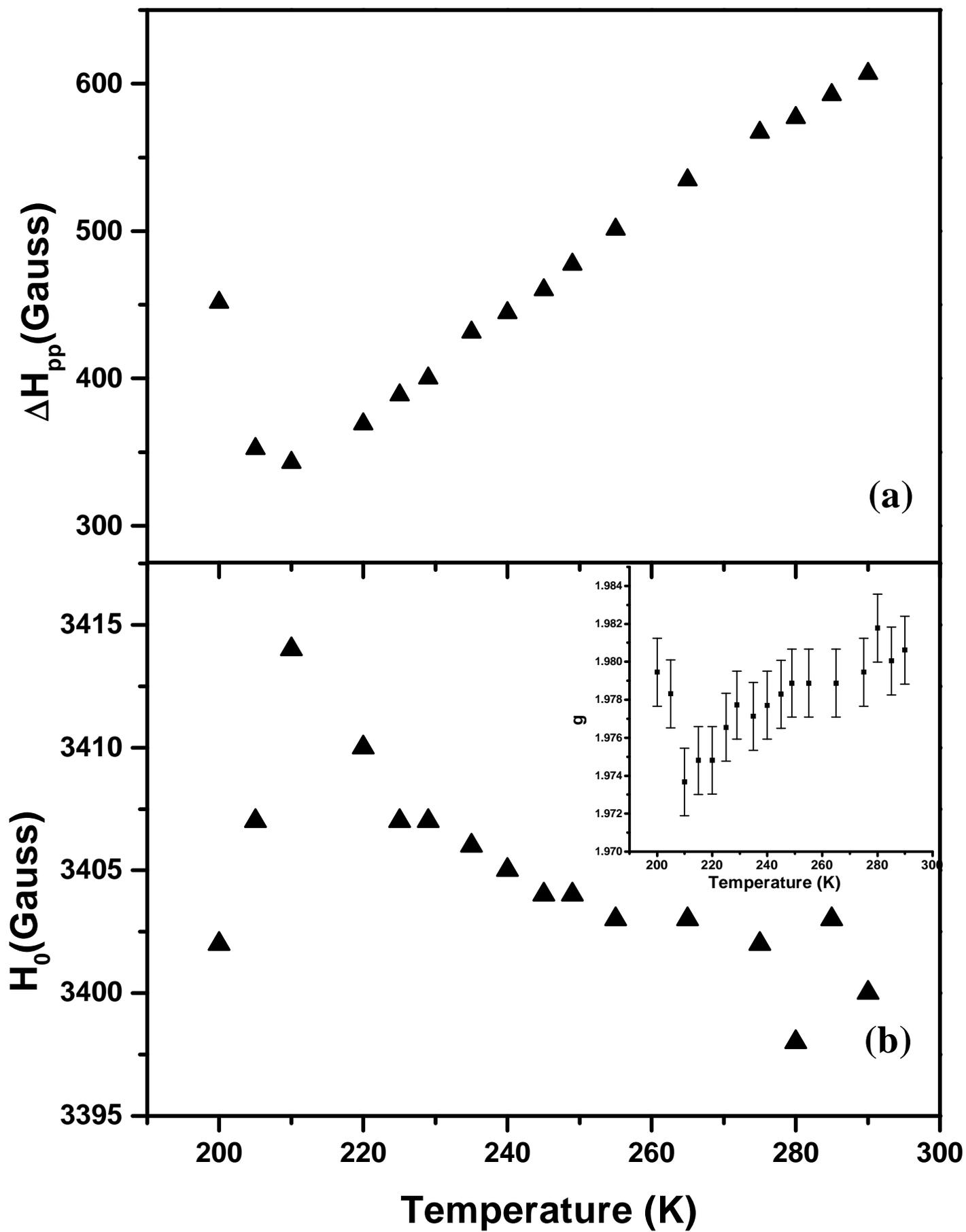
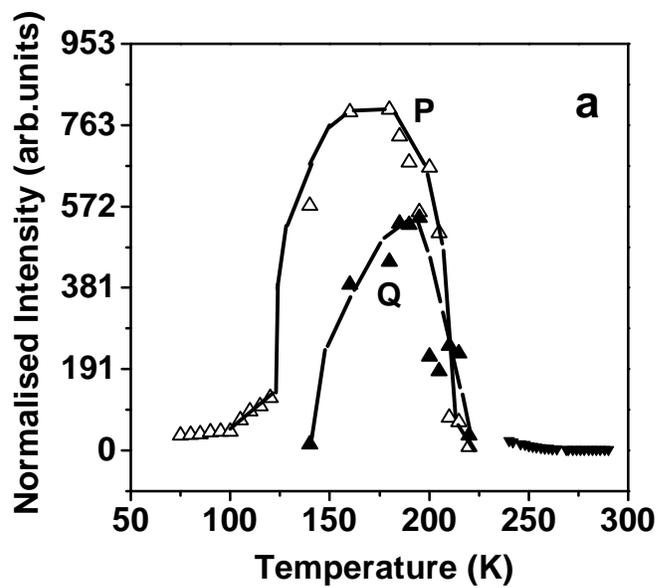
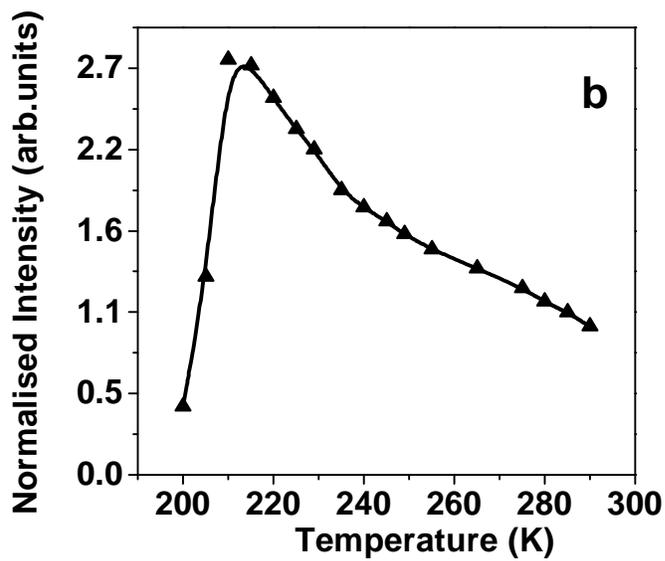
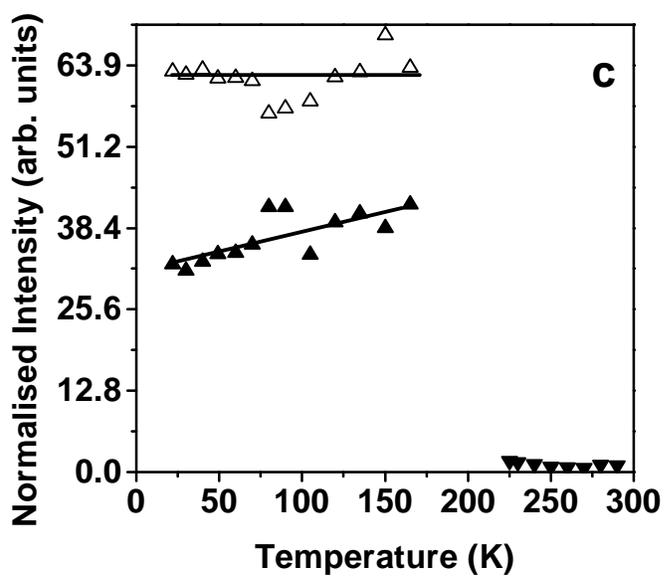
\end{document}